# Characterization of aromaticity in analogues of Titan's atmospheric aerosols with Two-Step Laser Desorption Ionization Mass Spectrometry

Ahmed Mahjoub[1*#], Martin Schwell[1], Nathalie Carrasco[2,3], Yves Benilan[1], Guy Cernogora[2], Cyril Szopa[2,3] and Marie-Claire Gazeau[1]

[1] LISA UMR CNRS 7583, Université Paris Est Créteil and Université Paris Diderot, Institut Pierre Simon Laplace, 61 Avenue du Général de Gaulle, 94010 Créteil, France

[2] Université Versailles St-Quentin, UPMC Univ. Paris 06, CNRS/INSU, LATMOS-IPSL, 11 Bd d'Alembert, 78280 Guyancourt, France.

[3] Institut Universitaire de France.

*Corresponding Author: ahmed.mahjoub@lisa.u-pec.fr

**Proposed Running Head**: aromatic compounds in analogues of Titan's aerosols

**Editorial correspondence to:**

Dr. Mahjoub, LISA UMR CNRS 7583, Université Paris Est Créteil and Université Paris Diderot, Institut Pierre Simon Laplace, 61 Avenue du Général de Gaulle, 94010 Créteil, France.

# Actuel adresse : Jet Propulsion Laboratory, California institut of technology, Pasadena, California, USA.

E-mail address: ahmed.mahjoub@latmos.ipsl.fr

**Abstract:**

The role of polycyclic aromatic hydrocarbons (PAH) and Nitrogen containing PAH (PANH) as intermediates of aerosol production in the atmosphere of Titan has been a subject of controversy for a long time. An analysis of the atmospheric emission band observed by the Visible and Infrared Mapping Spectrometer (VIMS) at 3.28 µm suggests the presence of neutral polycyclic aromatic species in the upper atmosphere of Titan. These molecules are seen as the counter part of negative and positive aromatics ions suspected by the Plasma Spectrometer onboard the Cassini spacecraft, but the low resolution of the instrument hinders any molecular speciation.

In this work we investigate the specific aromatic content of Titan's atmospheric aerosols through laboratory simulations. We report here the selective detection of aromatic compounds in tholins, Titan's aerosol analogues, produced with a capacitively coupled plasma in a $N_2$:$CH_4$ 95:5 gas mixture. For this purpose, Two-Step Laser Desorption Ionization Time-of-Flight Mass Spectrometry (L2DI-TOF-MS) technique is used to analyze the so produced analogues. This analytical technique is based on the ionization of molecules by Resonance Enhanced Multi-Photon Ionization (REMPI) using a λ = 248 nm wavelength laser which is selective for aromatic species. This allows for the selective identification of compounds having at least one aromatic ring. Our experiments show that tholins contain a trace amount of small PAHs with one to three aromatic rings. Nitrogen containing PAHs (PANHs) are also detected as constituents of tholins. Molecules relevant to astrobiology are detected as is the case of the substituted DNA base adenine.

**Introduction:**

Titan is the only known satellite in the solar system with a dense atmosphere. This atmosphere is even more substantial than the Earth atmosphere, with a pressure of $1.5 \times 10^5$ Pa at the surface. Dinitrogen ($N_2$) and methane ($CH_4$) are the predominant species in Titan's atmosphere. Energetic particles coming from Saturn's magnetosphere and/or solar UV photons bring appropriate energy to dissociate both $N_2$ and $CH_4$ into radicals and ions. The latter trigger a complex organic chemistry in the upper atmosphere, based on numerous C, H and N compounds (Waite et al., 2007). This complex chemistry leads to the formation of aerosols layers thick enough to drive the radiative balance of the satellite. These aerosols were shown to be composed of a solid organic part made of organic oligomers bearing C, H and N atoms (Israel et al., 2005; Waite et al., 2007). While the composition of this material and the way it is produced are far from being fully characterized, it is more and more believed that macromolecules such as HCN polymers, $C_2H_2$ polymers and aromatic molecules, including polycyclic aromatic hydrocarbons (PAH), should play important roles in the transition from the gas phase to solid particles (Wilson & Atreya, 2003). Especially molecules containing aromatic rings have been suspected for a long time to be involved in the mechanisms of aerosol growth. This hypothesis, as well as the detection of benzene in the upper atmosphere of Titan (Waite et al., 2007, Vinatier et al., 2010), enhanced development of models studies explaining the formation of aromatics in Titan's atmosphere. These models predict the formation of PAHs as a first step of the formation of Titan's haze (Wilson & Atreya, 2003; Lebonnois et al., 2002). Moreover, limb daytime observations of Titan's upper atmosphere with the Visual Infrared Mapping Spectrometer (VIMS) onboard the Cassini spacecraft show a strong emission band around λ = 3.28 µm (Dinelli et al., 2013). This band is attributed to the emission of aromatic compounds (Lopez-Puertas et al., 2013). Today it is widely believed that aromatic molecules play a major role in this chemistry as intermediates between small molecules and polymers big enough to condensate as aerosols in Titan's atmosphere. The Ion Neutral Mass Spectrometer (INMS) on board Cassini is capable to detect molecules with *m/z* up to 99, however its low mass resolution hinders the identification of molecules.

Limitations of the *in situ* measurements of Titan's aerosols make it necessary to develop theoretical and experimental simulations in order to investigate Titan's atmospheric chemistry. One powerful methodology is to synthesize and study the physical chemistry of analogues of Titan's aerosols (so called tholins) in the laboratory. In the past, photochemical reactors or plasma discharge experiments have been used for this purpose (Khare et al., 1984; Coll et al., 1999; Tran et al., 2003; Imanaka et al., 2004; Szopa et al., 2006; Vuitton et al., 2009; Hasenkopf et al., 2010, Carrasco et al., 2013). Among them, plasma discharge reactors are the most efficient ones in order to produce large amounts of tholins.

The formation of PAHs and PANHs has been particularly considered by many teams because of the general importance of these molecules in astrophysics and also because of their relevance in prebiotic chemistry. Khare et al. (1984) carried out pyrolysis Gas Chromatography-Mass Spectrometry (GC-MC) experiments with tholins and reported the detection of aromatic molecules and N containing aromatics in the pyrolysis products. Ehrenfreund et al. (1995) pyrolyzed tholins produced by bombardment of a $N_2/CH_4$ gas mixture with electrons. Measurements were carried out by Curie-point-GC-MS and by temperature-resolved in-source pyrolysis-MS. They observed benzene, toluene and pyridine. Hodyss et al. (2004) recorded the 3-D-fluorescence spectra of tholins in acetonitrile and water. They detected an emission band at 397 nm. They mentioned that PAH's could be responsible of this band while linear conjugated systems have also similar emission. Excellent discussions about the detection of aromatics in laboratory synthesized tholins can be found in review papers dating before and after Cassini Era (Coll et al., 1998 Cable et al., 2012). Derenne et al. (2012) reported the detection by solid NMR of heterocyclic aromatic species containing $C_3N_3$ and $C_6N_7$ units as well as carbons involved in imino groups. This observation is in agreement with infrared spectra of tholins produced by the same reactor (PAMPRE) reported by Gautier et al. (2012). To investigate the role of benzene on the chemistry of Titan's aerosols, Trainer et al. (2013) studied tholins produced in photochemical reactor in $N_2$-$CH_4$ gas mixture containing trace amounts of benzene. They showed that the introduction of few tens of ppm of benzene can have a significant effect on the composition of the produced tholins. Authors proposed a photochemical reaction between benzene and methane as a pathway to produce larger PAH such as naphthalene. The same group reported Far-IR spectra of tholins produced with traces of aromatics and N containing aromatics (Sebree et al., 2014). They observe a strong band at 200 cm$^{-1}$, which they tentatively assign to PANH's.

The aim of the present study is to identify univocally aromatic molecules in tholins with a specific method described in detail below. For tholins synthesis we use the reactor described by Szopa et al. (2006) which is a plasma RF discharge applied to a $N_2$/$CH_4$ gas mixture. The chemical and physical properties of tholins produced by this set-up have been studied previously by a variety of techniques namely infrared spectroscopy (Quirico et al., 2008; Gautier et al., 2012), elemental analysis (Sciamma-O'Brien et al., 2010), mass spectrometry (Pernot et al., 2010; Carrasco et al., 2009), solid nuclear magnetic resonance (Derenne et al., 2012) and ellipsometry (Sciamma-O'Brien et al., 2012; Mahjoub et al., 2012; Mahjoub et al., 2014). Some of the previously reported results using non selective techniques give hints on the possibility of presence of aromatics in tholins. These results cannot be considered as a detection of aromatics since other species can also explain them (for example infrared bands assigned to aromatics can be also due to linear conjugated molecules). In this work, we present a selective detection of aromatic molecules based on an analytical mass spectrometry technique: (L2DI-MS). Using such a discriminating technique allows us to examine the aromaticity of aerosols without interference from non-aromatic species. The benefit of L2DI-MS technique compared to previously used analytical methods is the ability to probe the aromatic part of aerosols with no ambiguity. Such measurements are necessary to address with more confidence the long debated question of aromaticity of tholins.

**Experimental details:**

1- Production of tholins

The experimental setup dedicated to the production of Titan's aerosol analogues is based on a Capacitively Coupled Plasma Radio Frequency (CCP RF) operating at 13.56 MHz (Szopa et al., 2006; Alcouffe et al., 2010). The reactor is a stainless steel cylinder 30 cm in diameter and 40 cm high. The discharge is applied between a polarized electrode and a grounded cylindrical grid box of 13.7 cm confining the plasma. The gas mixture (95 % $N_2$ and 5 % $CH_4$) flows continuously through the shower shape polarized electrode with a rate of 55 standard cubic centimeter per minute (sccm). It is pumped by a rotary pump and the pressure in the reactor kept constant to approximately 0.9 mbar. Before sample production, the reactor is heated and pumped down to a secondary vacuum by a turbomolecular pump. When the discharge is on, the RF power generator delivers 30 W. In these conditions the dusty plasma produces solid organic material as brownish-orange powders (Sciamma-O'Brien et al., 2010). These powders grow inside the plasma discharge while being in levitation between the electrodes. Above a critical size they are ejected from the plasma by the gas flow and collected inside a surrounding glass vessel. After some time, the experiment is stopped and the powder can be taken out of the vacuum chamber for further investigation. This experimental set-up is named "PAMPRE" which is the French acronym for *Prodcution d'Aérosols en Microgravité par Plasmas Réactifs.*

2- Resuspension of tholins powder prior to analysis by L2DI-MS

In order to introduce a sample of tholins in our L2DI instrument, 1.5 g of tholins particles are put into water at a ratio of approximately 3 g.L$^{-1}$. This aqueous mixture is then ultrasonicated. During this procedure part of the tholins are dissolved whereas the water-insoluble fraction stays in suspension. While being stirred constantly in order to avoid sedimentation of the non-soluble fraction, the mixture is nebulized using a commercially available constant output atomizer (model 3076 from TSI) using 1.5 bar of $N_2$ as a carrier gas prior to expansion through a special nozzle. This procedure permits to produce an aqueous aerosol in the gas phase at atmospheric pressure that contains both the water-soluble and the water-insoluble fractions. Water was chosen as nebulizing liquid for two reasons: 1) it permits optimal working conditions of the atomizer and 2) it is known that tholins are more soluble in polar solvents (Carrasco et al., 2009).

Hydrolysis of the soluble fraction in water is a point debated in several studies (Khare et al., 2001; Raulin et al., 2007; Neish et al., 2008 & 2010, Poch et al., 2012; Cleaves et al., 2014). It can produce prebiotic molecules of high interest in astrobiology. But the kinetics is slow, so that it is often enhanced in the laboratory by warm, acidic or alkaline conditions. Raulin et al (2007) showed that it occurs after 2 days at 70°C. And finally, even in harsh acidic conditions, Khare et al. (2001) showed that only 1% of amino acids was produced by mass of tholins. Knowing that our samples are few soluble and that the colloidal suspension is prepared in water only during the few hours duration of the experiments and at room temperature, we expect the issue of the hydrolysis to be non-critical in our case.

Downstream the atomizer nozzle, a commercially available diffusion dryer (model 3062 from TSI) is used to remove the water from the aerosol. This provides an aerosol that contains mainly dry solid particles and its composition is thought to be very close to the initially plasma-produced tholins. The size distribution of the so-produced aerosols will be given in a forthcoming publication.

3 - Laser Two-Step Desorption Ionization - Mass Spectrometry (L2DI-MS) of tholins

The chemical analysis of tholins is done by a laser aerosol mass spectrometer that has been developed and optimized before in the LISA laboratory. This instrument (initially called "SPLAM" for Single Particle Laser Ablation Mass Spectrometer) has been described in details by Gaie-Levrel et al. (2012) and only a brief description will be given here. Figure 1 gives a scheme of the instrument. Briefly, the particles are spatially focused into the ionization chamber using an aerodynamic lens system (ALS). These kinds of nanoparticle lenses have been developed initially by the atmospheric physics community and are widely used for aerosol sampling. Along the ALS, the pressure drops from 1 bar (at the inlet) to about $10^{-3}$ mbar in the first vacuum chamber pumped by a 250 l/s turbomolecular pump (TMP). In all, 5 TMP are connected to the instrument for the three differential pumping steps and the mass spectrometer. In the first vacuum chamber the particles are already focused to a beam of very low divergence and a diameter of less than 1 mm. They continue to travel along a straight-line path through a second chamber to arrive in the detection/sizing chamber with a vacuum of approximately $10^{-5}$ mbar. Here, the particle beam crosses two cw diode lasers ($\lambda$ = 403 nm) with a power of about 30 mW each. This chamber is blackened inside and the light diffusion of particles can be easily observed by eye or, optionally, by photomultipliers. This light detection is important since it permits the verification of the proper functioning of the whole aerosol inlet. The use of two lasers permits sizing of particles by measuring their flight time between the two crossing points (in the case of low number density of the introduced aerosol). However, in this study, the sizing system has not been used to further investigate the size distribution of the introduced aerosol since we focus on the chemical composition of tholins. For this reason high number densities of particles have been introduced into the instrument which is incompatible with a size measurement.

Following the sizing chamber, the particle beam enters into the ionization chamber which is pumped down to about $10^{-7}$ mbar. In the center of this chamber, pulsed laser radiation from two lasers is used to respectively 1) desorb matter from the particles and 2) ionize the desorbed molecules. The two steps are separated in time and space taking into account the travel speed of the particles (about 140 m/s). The time delay (about 10 µs) between the two lasers is optimized by maximizing the L2DI mass signal of some typical ion. This procedure is also used for the spatial gap of the two lasers (~2 mm). Note that in the previous configuration of the instrument (cf. Gaie-Levrel et al., 2012) only one UV laser (248 nm) has been used to desorb and ionize single particles in one step (this process is generally called "LDI" for Laser Desorption Ionization). Here, we use a pulsed $CO_2$ IR laser ($\lambda$=10.6 µm) from ALLTEC delivering up to 1 J of pulse energy. The initial cross section of the laser beam is 15x18 mm and the light is homogenously distributed in this rectangular area. The beam is focused by a ZnSe lens to about 1 mm$^2$ beam size at the particle interaction zone. Given the pulse length of about 10 µs, the IR power density in the interaction region is about $10^7$ W.cm$^{-2}$. According to Woods et al. (2002) this is sufficient to desorb solid particles containing organic molecules. The power density can be adjusted continuously by crossed ZnSe polarizers as well as by changing of focusing conditions.

Vaporized molecules are ionized by UV laser radiation from a KrF excimer laser emitting at $\lambda$ = 248 nm (Optex, Lambdaphysik). This laser is capable of delivering up to 15 mJ of pulse energy inside its 10x20mm beam cross section. The actual applied pulse energy

(regulated also by crossed polarizers and focusing conditions) is, however, much lower and adjusted in order to ionize aromatic molecules softly by Resonant Two-photon ionization (R2PI) (see below).

Both lasers are synchronized and operate at a repetition rate of 5 Hz. Ions formed are detected with a home-made Wiley-McLaren Time-of-Flight Mass Spectrometer (TOF-MS) with 1 m field free drift region. The acceleration potentials used are +145 and +1500 V respectively for primary extraction and subsequent ion acceleration. These conditions yield a mass resolution of about m/Δm = 300 at *m/z* 92 as determined by recording a mass spectrum of gaseous toluene admitted from an effusive inlet to the ionization chamber.

The L2DI-MS technique has been applied before to the mass spectrometry of aerosols. For example an IR/VUV laser radiation scheme has been used in the group of T. Baer in the early 2000 years (Woods et al., 2001 & 2002, Smith et al., 2002). When VUV laser radiation is used for the ionization step (for example at λ = 118 nm like in Woods et al., 2001 & 2002, Smith et al., 2002) organic molecules, aliphatic or aromatic, are ionized with one photon ("single photon ionization") very close to the ionization threshold. Under these conditions dissociative ionization pathways are of minor importance (if the parent ion is stable) and mass spectra are essentially fragment-free. Later on an IR (λ = 10.6 µm)/UV scheme has been used for L2DI chemical analysis of real-world aerosols by Bente et al. (2008). These experiments are very close to ours since Bente et al. (2008) also use a 248 nm KrF excimer laser for (one-color) Resonance-Enhanced Multiphoton Ionization (REMPI) following IR laser desorption. They have shown in their work that this technique is "very suitable" (meaning quasi-selective) for PAHs and, additionally, that the number of UV-photons absorbed can be controlled efficiently (to 2 or 3) in order to minimize fragmentation. REMPI and R2PI need molecules that have a high absorption cross section (σ) at the UV laser wavelength. This is the case for aromatic molecules such as PAHs and PANHs at 248 nm where intense $S_0 \rightarrow S_n$ transitions are located. Their absorption cross sections range from $10^{-18}$ to $10^{-16}$ cm². In Table 1 are compared the absorption cross section of examples of non-aromatic molecules with those of examples of aromatics.

During our experiments the laser pulse energy is kept very low in order to avoid absorption of multiple photons. These conditions have been verified continuously by recording mass spectra of gaseous toluene: the laser pulse energy is adjusted in order to observe almost exclusively the parent ion *m/z* 92. We also cite the work of Carpentier et al. (2013) on the R2PI of PAHs generated by ethylene flames. Very much like Bente et al. (2008) they show that REMPI is selective for PAHs and that fragmentation can be reduced efficiently. In both experiments (Bente et al., 2008; Carpentier et al., 2013) only aromatics are observed although the chemical mixtures under study contain most certainly non-aromatic molecules. The interested reader is also referred to earlier work of Lustig & Lubmann (1991) addressing the question of selective ionization of aromatics by REMPI and R2PI, and to Vertes and Gijbels (1993) for fundamental work on matrix assisted laser desorption and subsequent laser ionization.

We finally cite the work of Hoon Hahn et al. (1988) and Sagan et al. (1993) who use L2DI-TOF-MS for the analysis of carbonaceous chondrites fallen on Earth and Tholins, respectively. In these studies, macroscopic samples have been placed directly in the ion source of the mass spectrometer. We will compare our results to those of Sagan et al. (1993). We note that the ultimate goal of our experiment is to directly sample the freshly formed tholins from the PAMPRE reactor in order to study also the dynamics of aerosol formation. This has not been achieved yet but will be done in the future.

The sensitivity of our REMPI-TOF-MS instrument has been characterized by direct injection of a variety of gaseous aromatic molecules into the ionization chamber (Gaie-Levrel, 2009). From these experiments it has been concluded that typically a few hundred molecules are needed inside the ionization volume (i.e. the cylinder formed by approximately 1 cm of the UV laser beam that is seen by the TOF-MS primary extracting field) to give a detectable mass signal when about 100 mass spectra are averaged.

|  | Non aromatics | | | | | |
|---|---|---|---|---|---|---|
| Molecule | α-pinene (a) | pentadiene (b) | butadiyne (c) | 1,2,3-Triazole (d) | pyrazole (e) | cyclohexane (f) |
| absorption cross section around $\lambda = 248$ nm (cm$^2$) | $1.0 \times 10^{-20}$ | $5.24 \times 10^{-21}$ | $7 \times 10^{-20}$ | $5.6 \times 10^{-20}$ | $-5.5 \times 10^{-20}$ | $1.21 \times 10^{-19}$ (at 179 nm) |
|  | Aromatics | | | | | |
| Molecule | benzene (g) | naphtalene (h) | pyridine (i) | pyrene (j) | anthracene (h) | dimethylaniline (k) |
| absorption cross section around $\lambda = 248$ nm (cm$^2$) | $2.9 \times 10^{-18}$ | $2.3 \times 10^{-17}$ | $0.6 \times 10^{-17}$ | $4.34 \times 10^{-17}$ | $2.3 \times 10^{-17}$ | $3.8 \times 10^{-17}$ |

Table1: Comparison between UV absorption cross section of examples of small nitriles and aromatic molecules.

a : Kubala et al., 2009,; b: Serio et al., 2012; c : Fahr & Nayak et al., 1994 d: Plamer et al., 2011; e : Walker et al., 2003; f : Pickett et al., 1951 ; g: Hiraya & shobatake, 1991 ; h: Ferguson et al. 1957; i : Bolovinos et al. 1984; j: Thony & Rossi et al. 1997; k: Kimura et al. 1964

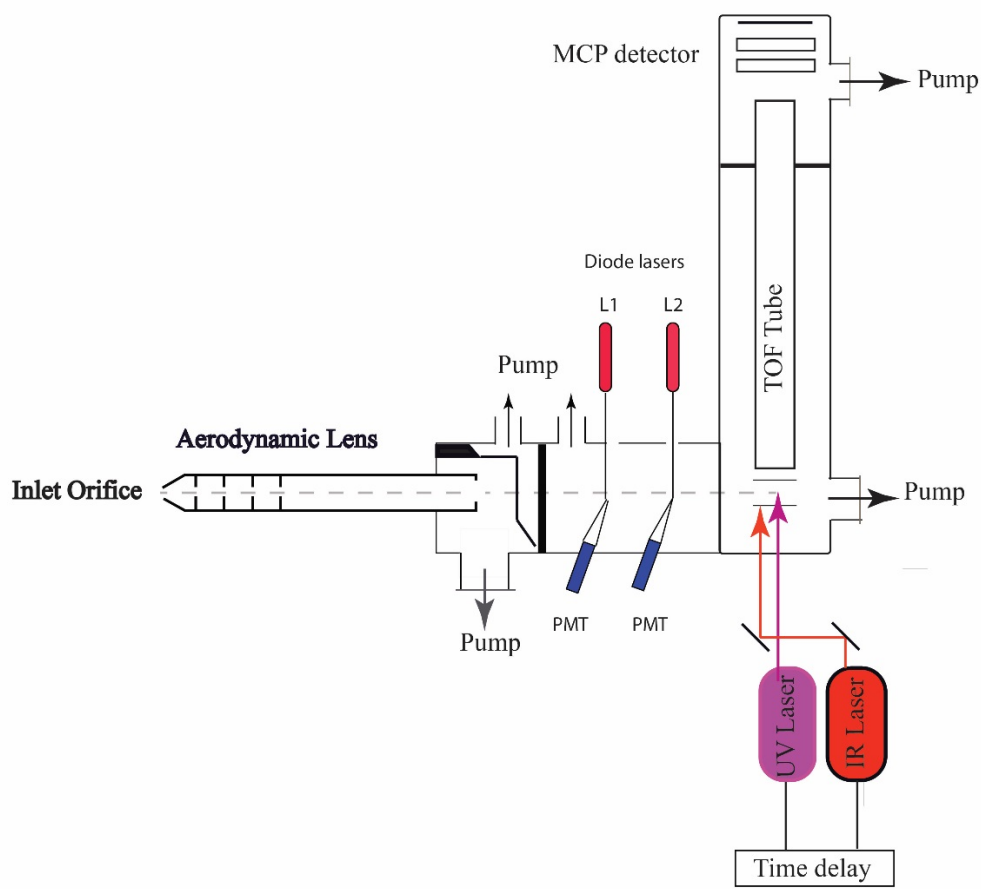

**Figure 1:** Schematic diagram of Laser 2 steps Desorption Ionization Mass Spectometry (L2DI-MS) technique (figure updated from Gaie-levrel et al. 2012).

## Results and discussion

1. <u>L2DI-MS of Dioctylphtalate (DOP) particles: validation of the technique</u>

As a preliminary test to validate our instrument we record a L2DI-MS spectrum of Dioctylphtalate (DOP) (Sigma Aldrich, >99 % purity). DOP ($C_{24}H_{38}O_4$; M = 390.55 g/mol) is a viscous oil with very low vapor pressure. The pure substance is easily nebulized with our TSI constant output atomizer. The low volatility of DOP is obviously important if one wants to maintain the DOP particles at an appreciable size while travelling through the vacuum of the instrument. Furthermore, DOP is an aromatic molecule and is therefore a good test substance for our L2DI instrument since it can be ionized with R2PI.

Figure 2 presents three DOP mass spectra. Fig. 2a is the one that can be found in the NIST Mass Spec Data Center. These MS are generally obtained with electron impact (EI) as ionization method, using electrons with 70 eV kinetic energy. The ions observed in the NIST DOP MS are: *m/z* 279 ($C_{16}H_{23}O_4^+$); *m/z* 167 ($C_8H_7O_4^+$); *m/z* 149 ($C_8H_5O_3^+$); m/z 113 ($C_8H_{17}^+$); m/z 112 ($C_8H_{16}^+$) and m/z 83 ($C_6H_{11}^+$). Note that the DOP parent ion (m/z 390) is not observed. Fig. 2b shows our L2DI-MS mass spectrum averaged over about 100 spectra. Obviously, mass spectra 2a (NIST) and 2b (our work, L2DI) are very similar. This is astonishing since EI ionization with 70 eV electrons usually gives rise to much more fragmentation than photoionization close to the threshold. However, this can be understood considering the low stability of the parent ion of DOP that is hardly observed even with single photon ionization close to its ionization threshold (at ~8.2 eV) as is evident from experiments with tunable VUV radiation conducted by Gaie-Levrel et al. (2011) (please cf. figure 4 in Gaie-Levrel et al. 2011). Close to the ionization energy (IE) of DOP, only the fragment ion *m/z* 279 is observed formed obviously by loss of one of the aliphatic side chains, $C_8H_{15}$. It is known that aliphatic side chains can be lost easily from aromatics in mass spectrometry since the resulting aromatic cations are quite stable. For our study, this means that the analysis of tholins mass spectra shall not consider parent ions of aromatic molecules with aliphatic side chains as observable by L2DI (R2PI).

It is important to note here that, on the contrary, the parent ions of non-functionalized PAHs (and by analogy probably PANHs) are stable enough to be observed by photoionization mass spectrometry even when excited with several electronvolts above their IE (cf. Jochims et al., 1997 and 1999, as well as references cited there). This explains that R2PI mass spectra of these compounds contain few fragments. The dissociative ionization pathway lowest in energy is always the H-loss reaction (about 3-12 eV above IE for the 37 PAHs studied by Jochims et al. (1999)). This loss reaction conserves thereby the molecular ring skeleton. Notice also that this holds for methyl-substituted PAHs too (cf. Jochims et al., 1999) whose parent ions can thus be considered as observable by our tholins L2DI-MS analysis. We finally note that even with EI ionization (70 eV impact energy) the base peak of PAH mass spectra is always the parent ion (cf. NIST Mass Spec Data Center).

In Fig. 2c we also show, for comparison, the one-step one-shot LDI mass spectrum recorded by Gaie-Levrel et al. (2012). This spectrum is much noisier than our L2DI spectrum since it comes from only one DOP particle that is analyzed by the TOF-MS, without any averaging. The heavier ions $m/z$ 279 and $m/z$ 167 are lacking in the single particle LDI MS of Gaie-Levrel et al. (2012). This can be explained by the higher internal energy released to the parent cation after absorption of multiple UV photons. This is also evidenced by the fact that ions of lower mass ($m/z$ <120) are more intense than in figures 2a, b relative to the base peak at *m/z* 149 that is observed in all 3 spectra. As explained above, the LDI technique uses one UV laser and therefore needs higher power densities than L2DI in order to simultaneously desorb and ionize particles. In these conditions the multiple (i.e. n >> 3) absorption of UV photons cannot be avoided.

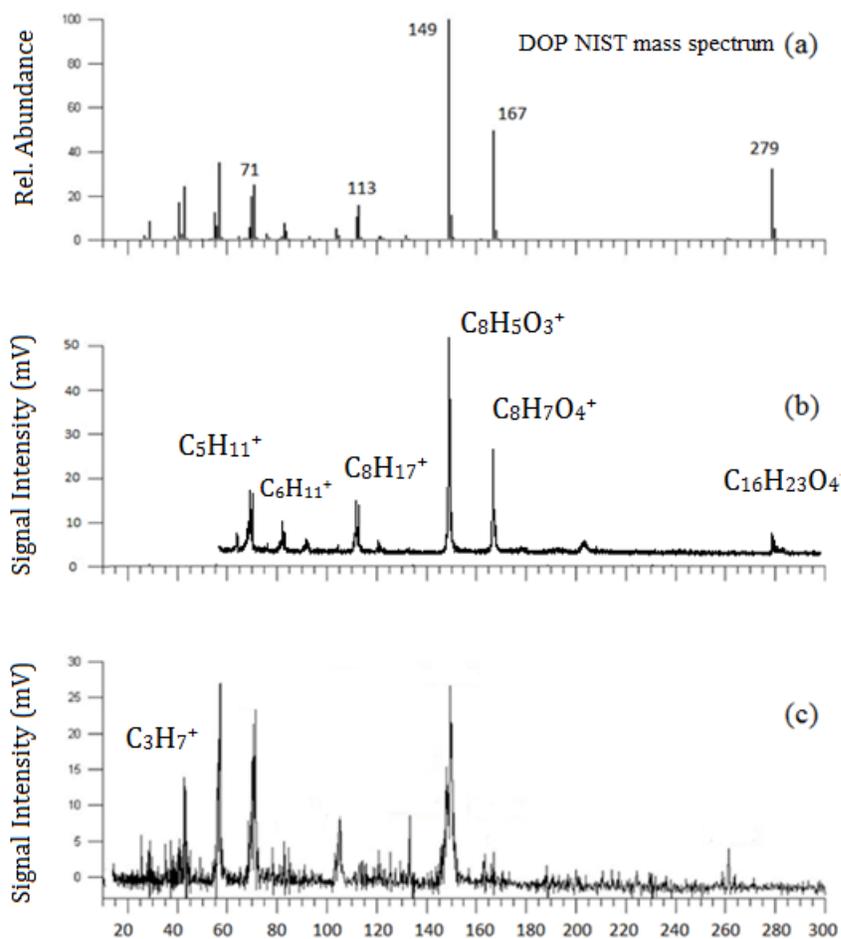

**Figure 2:** Dioctylphthalate ($C_{24}H_{38}O_4$) mass spectra a) Electron impact ionization mass spectrum (NIST 2010). b) L2DI-mass spectrum (this work), c) LDI single particle mass spectrum (Gaie-levrel et al., 2012)

2. <u>L2DI-MS mass spectrum of tholins</u>

The L2DI-MS technique relies on the desorption from particles by IR laser radiation and thus UV laser power density is kept low enough to avoid 1) absorption of multiple photons and consequently 2) fragmentation of PAHs and PANHs. Having a mass spectrum not affected by dissociative photoionization is of course decisive for the chemical analysis of tholins presented here. Indeed, the chemical components of tholins are not known very well and fragmentation would complicate the identification of the mixture of unknown compounds. Considering the literature presented above, we are, however, convinced that L2DI mass peaks observed here can be assigned to parent ions of PAHs, PANHs or their methylated derivatives.

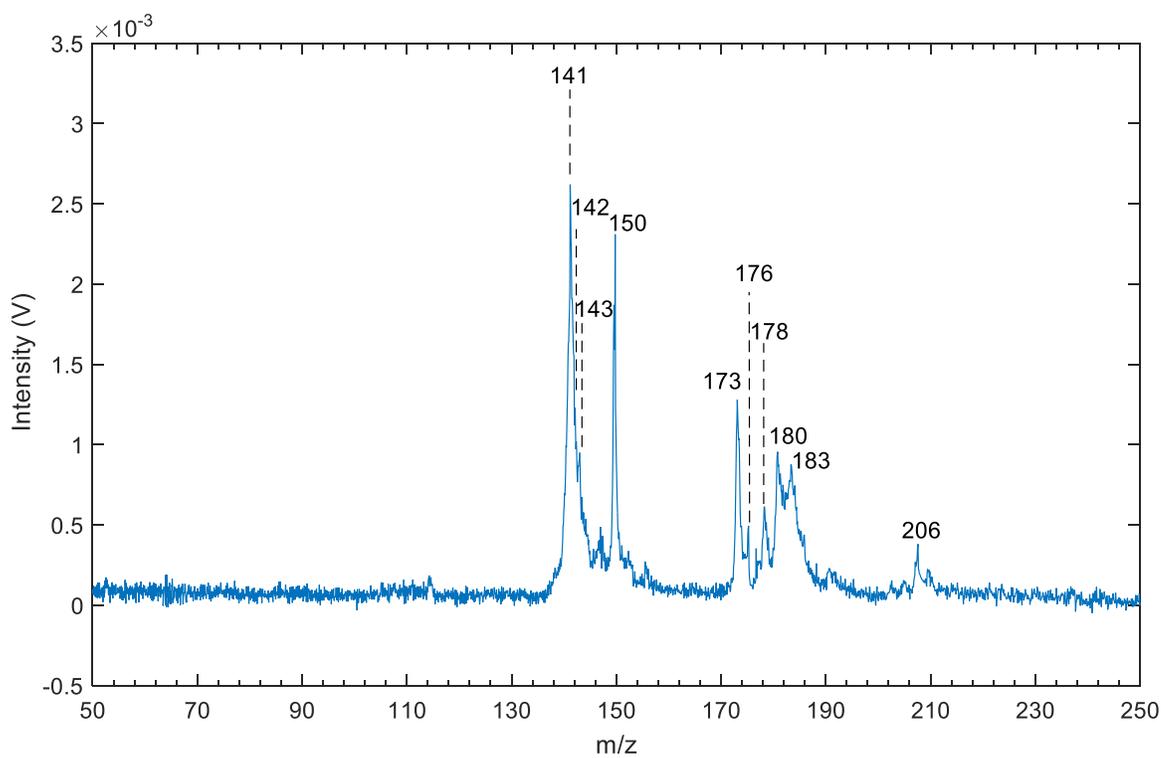

**Figure 3**: L2DI-MS spectrum of tholins produced from a $N_2/CH_4$ mixture (95%/5%).

Figure 3 presents an L2DI-MS of tholins particles. In this spectrum appear seven well resolved peaks at *m/z* 150, 173, 176, 178, 180, 183 and 206. Congested mass peaks are also observed between *m/z* 141 and 143 with a maximum at *m/z* 141. The *m/z* 1 to 50 as well as the *m/z* > 250 range do not present any observable peak, The absence of peaks below *m/z* 140 confirms that our technique induces little fragmentation through dissociative ionization. The L2DI-MS spectrum shows that a variety of aromatic molecules exists in the analyzed sample with ions detected in the *m/z* 141 to 206 range. The lack of detection of cations at below *m/z* 140 suggests also the absence of benzene and triazine in our sample. These two molecules have been detected in the gas phase of the PAMPRE reactor previously (Gautier et al., 2011). Pyrroles are apparently also missing in our mass spectrum. Using GCxGC-MS analysis, McGuigan et al. (2006) found that these molecules are among the most dominant species in tholins. A recent study by Morisson et al. (2016) shows that a possible pyrolysis temperature effect could be responsible for the production of these molecules. This could explain why no pyrroles are detected in our L2DI-MS spectrum. Another possibility for the explanation of the non-detection of benzene, triazine and pyrroles could be the evaporation of these semi-volatile compounds during the introduction of the tholins aerosols to our instrument.

Sagan et al. (1993) reported L2DI-MS spectrum for tholins produced by DC discharge in a $N_2/CH_4$ mixture. Sagan et al. detected mainly anthracene and penanthrene derivatives in Titan's tholins. The authors in this paper consider only PAHs in their assignment of the mass spectra and without giving the reason why they exclude the possibility of formation of PANHs variety of detected aromatics. In addition in our spectrum we detect molecules with *m/z* as small as 141, while the lower mass observed by Sagan et al. is 192 a.m.u. On the other hand, the larger molecule observed in our study is *m/z* 206 while Sagan et al observed a peak at 208 a.m.u.

First we can compare the L2DI-MS mass spectrum to the high resolution mass spectrum previously reported in Gautier et al. (2014) (Figure 4). These authors analyzed tholins produced also by the PAMPRE setup (in the same conditions than our samples) using Orbitrap High Resolution Mass Spectrometry (OHR-MS). For OHR-MS analysis, tholins were dissolved in methanol at a 1 mg mL$^{-1}$ concentration before they were infused in an ElectroSpray ionization source. The ions so formed were then analyzed using a high resolution mass spectrometer with a hybrid linear trap/orbitrap analyzer (for more details see Gautier et al., 2014). Figure 4-a shows the OHR-MS mass spectrum in the 120-260 mass range. We note that all mass peaks observed in the L2DI-MS spectrum are also present in the OHR-MS mass spectrum. Since the Orbitrap technique is not selective to aromatics, much more mass peaks are observed in the OHR-MS as compared to the L2DI-MS. On this first global comparison, we can deduce that tholins are mainly composed of non-aromatic organic species, but that a few aromatic compounds are firmly detected by L2DI-MS, with molecular masses between 140 and 206 a.m.u. We can also infer that species at *m/z* > 206 in the OHR-MS are probably not aromatics since these molecules are not detected in the L2DI spectrum.

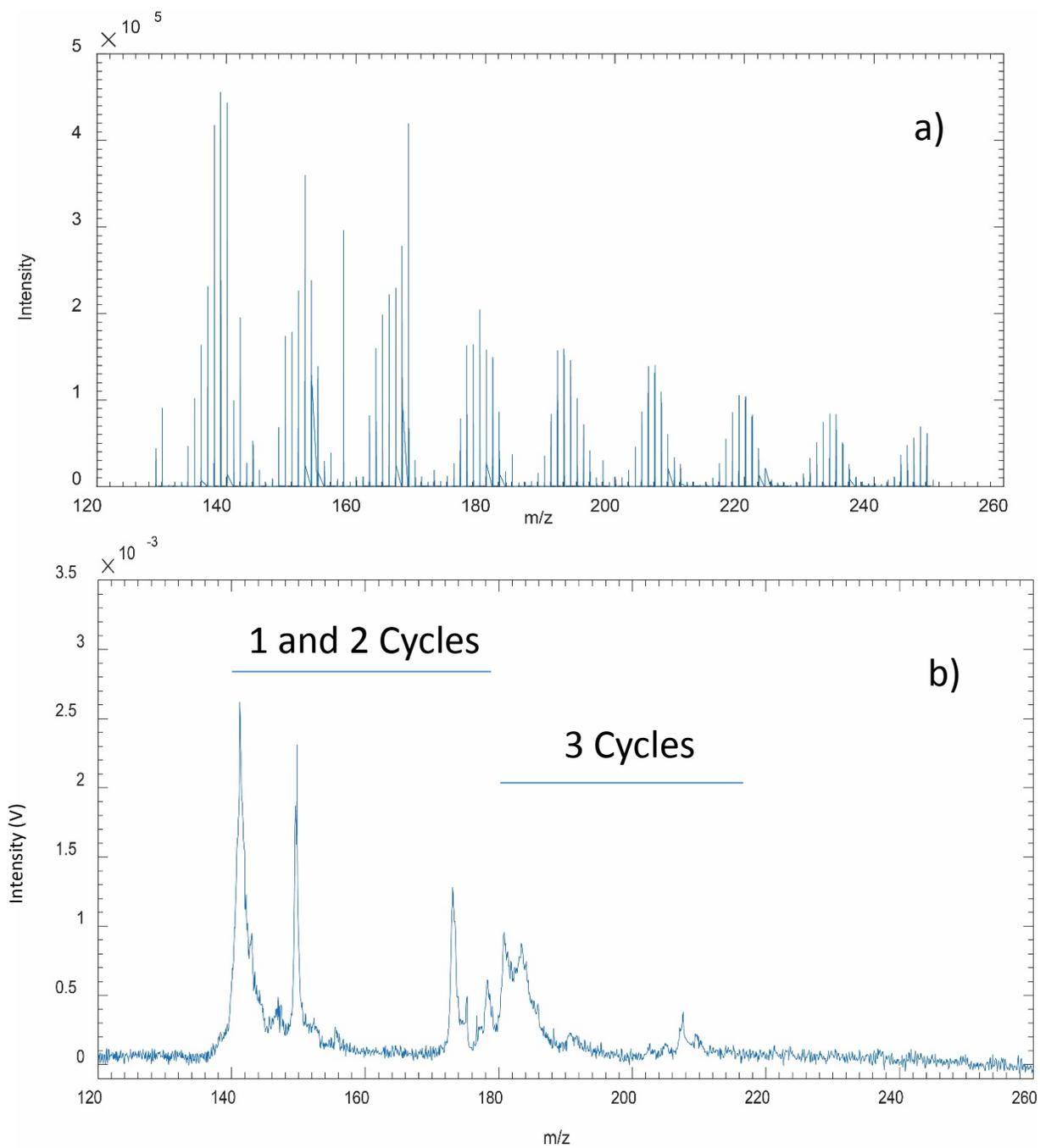

**Figure 4:** Comparison between: a) OHR-MS mass spectrum of tholins produced with a $N_2$-$CH_4$ gaseous mixture containing 5 % of methane (spectrum taken from Gautier et al. 2014). b) L2DI-MS spectrum of tholins produced in the same conditions.

**Possible assignments:**

Assignment approach:

The mass resolution of the TOF mass spectrometer used here is not high enough to infer directly the chemical composition of detected molecules. In order to assign mass peaks in the L2DI-MS, we use the chemical raw formula from the OHR mass spectra as a guide. The mass resolution of the OHR-MS is of about m/Δm = 200000 at $m/z$ 150. This opens up the possibility to attribute each observed mass peak to an ion with its chemical raw formula. The strategy is as follows:

- For each molecular species observed in the L2DI-MS, we take into account all the molecules listed in the NIST and Molpart (www.molpart.com) databases which have the same molecular weight and contain at least one aromatic ring. Since tholins are produced from a $N_2/CH_4$ mixture, we consider only molecules constituted by carbon, nitrogen and hydrogen atoms ($C_xN_yH_z$). By this assumption we do not take into account the possible oxidation of tholins during the exposure of the sample to air (Cable et al. 2012; Sciamma-O'Brien et al. 2010). We exclude molecules with aliphatic side chains larger than methyl.
- We compare the list of $C_xN_yH_z$ molecules so extracted for each mass to species corresponding to that mass detected by OHR-MS. Any aromatic species $C_xN_yH_z$ extracted in the first step from the NIST database and not detected by the OHR-MS technique is excluded from the list.

1- *m/z* 141:

A broad and intense mass peak is observed between $m/z$ 141 and $m/z$ 143. It corresponds also to the most intense peak in the ORH-MS spectrum in the $m/z$ 130-250 range. For this peak we will take into account all the aromatic species with *m/z* 141, 142, 143.

For *m/z* 141, the aromatic molecules listed in the NIST and Molport databases which produce an ion for *m/z* 141 are $C_{10}H_7N$ (pyrrolo (2,1,5-cd)indolizine , cyclobuto-quinoline, methanoquinoline, 1H-cyclopropa[h]quinolone and 1H-cyclobuta[de]quinolone) and $C_{11}H_9$( 2-naphtylmethyl radical) (table 2). The assignment to 2-naphtylmethyl radical is supported by the detection of methyl-naphthalene (see assignment of mass 142). Moreover, the activation energy of the H loss photo-dissociation of methylnaphthalene cation is around 2 eV (Fu-Shiuan, et al., 1990) while the ionization energy is only 7.91 eV, so after absorption of two 5 eV photons, the methylnaphthalene molecule gets enough excess energy to dissociate into 2-naphtylmethyl radical + H.

2- *m/z* 142:

In the NIST and Molport databases the molecular species with an aromatic ring and producing an ion for *m/z* 142 are listed in table 2. Two molecules could be assigned to this mass: methylnaphthalene, 1,4-Dihydro-1,4-methanonaphthalene.

The presence of naphthalene and its substituted derivatives in Titan's atmosphere as well as in laboratory analogues of Titan's aerosols are widely discussed in the literature. Sagan et al. (1993), reported the detection of substituted naphthalene in laboratory analogues of Titan's aerosols produced by plasma discharge. In a more recent study Sebree et al. (2010) have measured and calculated the dissociation energy of the H atom from the naphthalene to be as low as 168 Kcal/mol (7.2 eV). In the cited paper authors discussed the possibility of the formation of naphthalene radicals in Titan's atmosphere and its implication in the chemistry, particularly the formation of larger PAH's. Our results confirm the formation of naphthalene and its derivatives in analogues of Titan's aerosols.

3- *m/z* 143

Aromatics molecules found in NIST database producing an ion at *m/z* 143 can be divided into three different bi-cyclic aromatics derivatives: isoquinoline, naphthalene and quinoline and one benzene derivative: benzene substituted by a pyrrole cycle (table 2). The naphthalamine molecule was recently detected in the upper atmosphere of Titan by analyzing of the 3.28 µm band of the VIMS data (Dinelli et al. 2013).

4- *m/z* 150

This mass corresponds to a multitude of aromatic species as listed in the NIST database. These species can be divided into 3 molecular families: substituted adenine ($C_5N_6H_6$), substituted pyrazine ($C_9N_2H_{14}$) and substituted benzene ($C_9N_2H_{14}$). These two later molecules are not detected by OHR-MS spectroscopy, then we can eliminate the assignment of these molecules to this mass peak. The 2-aminoadenine chemical formula ($C_5N_6H_6$) was detected with a high signal in the OHR-MS spectrum. Among these species, as only $C_5N_6H_6$ was detected by OHR-MRS, we can assign the mass peak at *m/z* 150 to the substituted adenine molecule (table 2). The Amino substituted adenine was previously detected in Titan's tholins produced by DC plasma discharge using pyrolysis GC-MS analysis (Khare et al., 1984 and Coll et al., 1998). The detection of substituted adenine is a very interesting result since it confirms the possibility of formation of nucleic acid bases in agreement with the detection of these kind of biomolecules (including adenine) in the PAMPRE aerosols by Horst et al. (2012).

5. *m/z* 173

This mass corresponds to 3 molecular families as listed in the NIST database: substituted benzene, quinazoline derivatives or benzo-quinolizine molecule (table 2). Substituted quinazoline species were detected by pyrolysis GC-MS in Titan's tholins produced by DC discharge (Khare et al., 1984).

6. *m/z* 176

A small peak appears at the mass 176. This mass can be assigned to 9, 10-Dehydrophenanthrene.

6. *m/z* 178

An intense peak is observed at *m/z* 178. The elemental chemical formula which corresponds to this mass peak in the Orbitrap analysis is $C_{14}H_{10}$. Aromatic molecules corresponding to this formula in the NIST or Molport databases are listed in table 2. Within these molecules are included 3-rings PAHs: phenanthrene and anthracene. This mass can also be assigned to 2-rings molecules: diphenylacetylene, 6b, 8a-Dihydrocyclobut[a]acenaphhylene and cyclohepta[de]naphthalene. The detection of heterocyclic analogues of phenanthrene and antheracene (see assignment of mass 180) supports the assignment of the *m/z* 178 mass peak to these two 3-ring PAH's.

7. *m/z* 180

The *m/z* 180 mass peak can be assigned to two nitrogen containing PAH's named phenanthroline and phenazine. These two hetero-polycyclic molecules corresponds to the heterocyclic analogues of anthracene and pheneanethrene detected at *m/z* 178. Many isomers are possible (with different positions of the nitrogen atom), in table 2 we present only examples of these isomers.

8. *m/z* 183

This mass peak can correspond to a multitude of aromatic species as listed in the NIST database. These species can be divided into 2 molecular families: $C_{11}N_3H_9$ and $C_{13}H_{13}N$. $C_{13}H_{13}N$ are not detected by OHR-MS spectroscopy, then we can exclude the assignment of these molecules to the mass peak at *m/z* 183. We assign this mass to aminoperimidine ($C_{11}N_3H_9$) which is detected in the OHR-MS study (table 2).

9. *m/z* 206

This mass peak can be assigned to dimethyl-anthracene and dimethyl-phenanthrene (table 2). The dimethyl-penanthrene was previously detected by Sagan et al. (1993) in tholins produced by DC discharge in $CH_4$-$N_2$ mixtures mimicking Titan's conditions. We note that since the *m/z* 206 mass peak is the heavier molecule observed in our L2DI-MS (figure 4), the largest PAH detected in our study is composed of 3 aromatic rings. This mass can also be assigned to 4,5,9,10-tetrahydropyrene (table 2).

| Mass | Orbi-trap mass and formula | Tentative assignments |
|---|---|---|
| 141 | C₁₀H₇N<br>m/z 141.1693 | 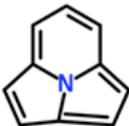pyrrole (2,1,5-cd)indolizine<br><br>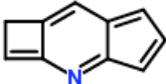cyclobuto-quinoline<br><br>Methanoquinoline 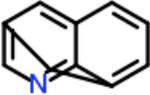<br><br>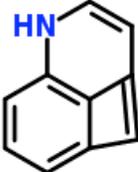1H-Cyclobuta[de]quinoline<br><br>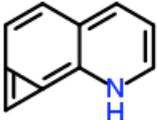1H-Cyclopropa[h]quinoline |
| | C$_{11}$H$_9$<br>*m/z 141.1892* | 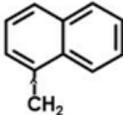<br>1-Naphthyl Methyl radical |
| 142 | C$_{11}$H$_{10}$<br>*m/z 142.1971* | 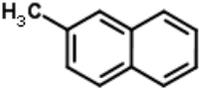 methylnaphtalene<br><br>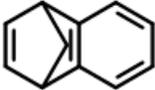<br>1,4-dihydro-1,4-methanonaphthalene |
| 143 | C$_{10}$NH$_9$<br>*m/z 143.1852* | 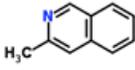 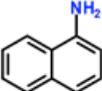 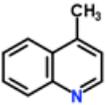 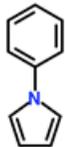<br>Methylisoquinoline   Naphthalamine   Methylquinoline<br>isoquinoline derivatives   substituted naphthalene   substituted Quinoline   Benzene derivatives<br>1H-Phenylpyrrole |
| 150 | C$_5$N$_6$H$_6$<br>*m/z 150.1413* | 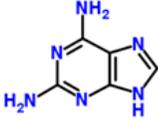<br>Substituted Adenine |
| 173 | C$_{10}$N$_3$H$_{11}$<br>*m/z 173.2144* | 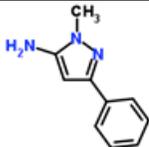<br>Substituted benzene<br><br>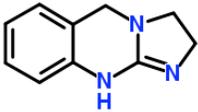<br>1,2,3,5-tetrahydroimidazo[2,1-b]quinazoline |
| | C$_{12}$NH$_{15}$<br>m/z 173.2542 | 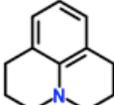<br>Benzo-quinolizine |
| 176 | C$_{14}$H$_8$<br>m/z 176.2133 | 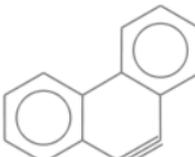<br>9,10-Dehydrophenanthrene |
| 178 | C$_{14}$H$_{10}$<br>m/z 178.2292 | 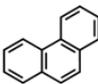 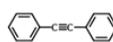 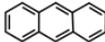<br>Phenanthrene    Diphenylacetylene    Anthracene<br><br>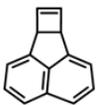    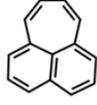<br>6b,8a-Dihydrocyclobut[a]acenaphthylene    Cyclohepta[de]naphthalene<br><br>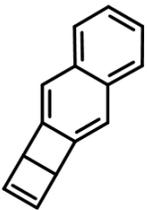<br>1,4-Dewar anthracene |
| 180 | C$_{12}$N$_2$H$_8$<br>m/z 180.2053 | 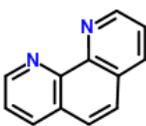    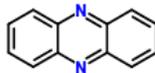<br>phenanthroline    Phenazine |

| 183 | C₁₁N₃H₉ m/z 183.2093 | 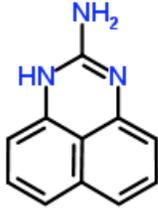 2-aminoperimidine |
|---|---|---|
| 206 | C₁₆H₁₄ m/z 206.2824 | 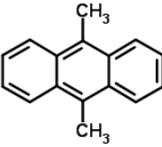 Dimethyl-anthracene    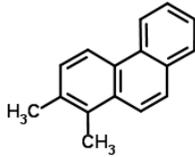 Dimethyl-phenanthrene 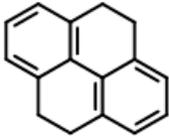 4,5,9,10-Tetrahydropyrene |

Table 2: Tentative assignment of mass peaks observed in the L2DI-MS spectrum of tholins

**Discussion:**

We report here the L2DI-MS analysis of Titan's aerosols analogues. The most crucial benefit from using this technique is its selectivity for aromatic molecules by using the REMPI ionization with a low UV laser pulse energy. The separation between desorption and ionization steps allows measuring mass spectra with little fragmentation.

Furthermore, Cabalo et al. (2000) reported the L2DI-MS spectra of aniline and compare it to mass spectrum of aniline without IR desorption (analysis of background gas). This study shows that the mass spectrum using IR laser to desorb molecules is comparable to the one of background gas. This indicate that using IR radiation to desorb molecules has no significant effect on the studied molecule.

This gives the L2DI-MS technique a major advantage compared to Pyrolysis GC-MS technique which is extensively used to characterize the aromaticity of tholins. Pyrolysis can add analytical artifacts with a synthesis of aromatics due to thermal processing (Morisson et al., 2016). L2DI-MS technique allows the detection of native aromatics representative of the original analyzed material.

The most important result deduced from the present study is the detection of two-ringed and three-ringed PAH's and PANH's as constituents of Titan's tholins produced by a Radio Frequency plasma discharge in nitrogen methane mixture. Peaks observed in the L2DI-MS spectrum are tentatively assigned to aromatic species based on a comparison between the molecular formula of aromatics existing in the NIST or Molport databases for mass peak and those given by OHR-MS spectroscopy. This comparison allows us to eliminate aromatic species present in these databases but not detected in OHR-MS.

The selective detection of aromatic species in tholins produced by PAMPRE experimental setup is a confirmation of previous indications of the existence of these species inferred by IR and NMR studies. Gautier et al. (2012) tentatively assigned infrared absorption bands observed at ~1600 cm$^{-1}$ in the absorption spectrum of tholins to aromatics and hetero-aromatics molecules. But these bands can also be due to C=C and C=N stretching modes. Solid NMR analysis of PAMPRE's tholins inferred the presence of heterocyclic aromatic species containing $C_3N_3$ and $C_6N_7$ units as well as carbons involved in imino groups (Derenne et al. 2012). In agreement with this NMR study, we assigned the majority of molecules observed in the L2DI-MS spectrum to heterocyclic aromatics and species containing a carbon involved in imino groups.

None of the techniques presented here (L2DI-MS and OHR-MS) is able to give a quantitative estimation of the proportion of aromatic molecules on the studied tholins. However, comparing the intensity and number of peaks of non aromatic species in the OHR-MS spectrum indicates that aromatic compounds are presents as trace in tholins. Sagan et al. (1993) used L2DI-MS technique to selectively detect PAH's constituents on Titan's tholins produced by DC discharge. They estimated the abundance of PAH's in their samples to be around $10^{-4}$ g.g$^{-1}$.

The detection of aromatics in Titan's aerosol analogues is a controversial subject: Coll et al. 1999 could not find any aromatic compounds except benzene in tholins produced by cold plasma discharge. However PAH's and PANH's were detected in Titan's tholins produced by other plasma discharge experiments (Khare et al. 1984; Sagan et al. 1993; Ehrenfrund et al. 1995 and Imanaka et al. 2004). Sagan et al. (1993) used L2DI-MS technique to characterize the aromaticity of tholins produced by DC-plasma discharge. The results obtained compare at same extend to our results but in their work they reported the detection of only PAH's and no nitrogen containing PAHs are attributed to any peak observed in their spectra. Hodyss et (2004) report the 3D fluorescence spectra of tholins produced by plasma discharge reactor and dissolved in water and acetonitrile, and then detected a fluorescence emission at 397 nm (in acetonitrile) and 305 nm (in water) similar to PAH's fluorescence spectra.

The formation of large PAH's and PANH's involve most likely the very reactive phenyl radical. The growth of PAH's can start through addition of two acetylene molecules on phenyl radical (Lebonnois et al.2002) leading to naphthalene ($C_{10}H_8$) (Wang et al. 1994 and Wong et al. 2000). Subsequent similar addition reactions will lead to higher PAH's. Addition of HCN or $HC_3N$ leads to the formation of heterocyclic PANH's (Ricca et al. 2001).

The benzene molecule is a key molecule for the formation of heavier PAH's as reported by all production scenarios of PAH's (e.g. Lebonnois et al. 2002). This molecule was not detected in our tholins as well as in tholins produced by DC discharge analyzed by L2DI-MS spectroscopy (Sagan et al. 1993). This molecule is volatile in room temperature and not expected to condense in solid tholins. Benzene has been detected in the gas phase of the PAMPRE experimental setup (Gautier et al., 2011) and is considered as progenitor of larger PAH's. It has also been detected in pyrolysis GC-MS analysis of Titan's aerosols analogues (see for instance Coll et al., 1998 and Khare et al., 1984). As mentioned above, thermal alteration during the analysis can be in the origin of synthesis of benzene observed after pyrolysis of tholins.

Another aromatic molecule has previously been detected in gas phase in PAMPRE experimental setup, namely triazine ($C_3H_3N_3$) (Gautier et al., 2011). This molecule has been tentatively detected in PAMPRE's tholins by NMR and RAMAN spectroscopy (Derenne et al., 2012; Quirico et al., 2008). In Quirico et al. (2008) triazine is detected only in tholins produced with a low concentration of $CH_4$ (2 %). In this paper, samples produced with 10 % concentration of methane do not show any signature of triazine. A recent study (Morisson et al. 2016) studied by pyrolysis-GCMS the chemical composition of tholins produced with different $CH_4$ concentrations (2, 5 and 10 %). triazine was detected exclusively in samples produced with 2 % of $CH_4$. All these studies indicate that triazine is only produced at low percentages of methane. This result is supported by the non-detection of this molecule and its derivatives in our L2DI-MS spectrum.

The presence of poly-aromatic molecules in Titan's aerosols has important consequences for Titan's atmosphere. These chromophores can absorb UV-Vis photons more efficient than other long Carbon and Nitrogen chains because of their delocalized π electrons. This absorbent character of PAH have major impacts on the optical properties of tholins in UV-Vis range (Imanaka et al. 2004). Moreover, a recent study shows that aromatic molecules have more ability to dissolve in liquid Methane compared to long hydrocarbon chains (Malaska et al. 2014). If PAH's and PANH's are present in Titan's hazes, these compounds have higher solubility in Titan's hydrocarbon lakes than linear hydrocarbons. This make them suitable target molecules for detection by Cassini-VIMS instrument or future mission visiting Titan's lakes. We suggest a better spectroscopic investigation of these compounds in liquid hydrocarbons.

**Conclusion:** We report here the characterization of aromaticity in Titan's tholins produced by RF electric discharge in $N_2/CH_4$ mixture using Two-Step Laser Desorption Ionization-Mass Spectroscopy $L^2$DI-MS. This technique is selective for aromatic molecules and this character allows us to reveal the aromatic part of tholins composition. Separating of desorption and ionization steps and using a low density of photons to ionize molecules permit the obtaining of a fragments free mass spectrum. Relatively small PAH's species (two to three rings) are detected in Titan's analogues produced by RF discharge. The lack of detection of aromatics larger than 3 cycles is in agreement with L2DI-MS aromatic analysis reported by Sagan et al. 1993. A tentative assignment of mass peaks observed in the L2DI-MS shows that both PAH and nitrogen containing PAH (PANH) are detected in the bulk tholins. Detected aromatic compounds can be subdivided into 6 families of molecules: benzene derivatives, naphthalene derivatives, indole derivatives, adenine derivatives, anthracene and it's N containing analogues, phenanthrene and its N containing analogues and bi-phenyl derivatives. In our L2DI-MS spectrum we did not detect benzene, triazine and pyrroles. The lack of detection of these molecules previously reported by different group using Pyrolysis techniques prompts the question of whether pyrolysis analyses represent the original composition embodying tholins. Certainly L2DI-MS technique does not have the ability to give isomer assignment of molecules detected but in the other side the selective detection of aromatics and the ability to analyze aerosols without much alteration make this technique suitable for aromaticity study of tholins.

**Acknowledgement:**

Dr. Mahjoub would like to thank Université Creteil Paris 12 for financial support. NC acknowledges the European Research Council for their financial support (ERC Starting Grant PRIMCHEM, grant agreement n°636829).

We thank the anonymous reviewer for his careful reading of our manuscript and many insightful comments and suggestions.